# Hybrid Fusion for Battery Degradation Diagnostics Using Minimal Real-World Data: Bridging Laboratory and Practical Applications

Yisheng Liu[a,b], Boru Zhou[a,b], Tengwei Pang[a,b], Guodong Fan[a,b,*] , Xi Zhang[a,b]

[a] School of Mechanical Engineering, Shanghai Jiao Tong University, 800 Dongchuan Road, Shanghai, 200240, China

[b] National Engineering Research Center of Automotive Power and Intelligent Control, Shanghai Jiao Tong University, Shanghai, 200240, China

*Corresponding author: guodong.fan@sjtu.edu.cn

## Abstract

Unpredictability of battery lifetime has been a key stumbling block to technology advancement of safety-critical systems such as electric vehicles and stationary energy storage systems. In this work, we present a novel hybrid fusion strategy that combines physics-based and data-driven approaches to accurately predict battery capacity. This strategy achieves an average estimation error of only 0.63% over the entire battery lifespan, utilizing merely 45 real-world data segments along with over 1.7 million simulated data segments derived from random partial charging cycles. By leveraging a thoroughly validated physics-based battery model, we extract typical aging patterns from laboratory aging data and extend them into a more comprehensive parameter space, encompassing diverse battery aging states in potential real-world applications while accounting for practical cell-to-cell variations. By bridging the gap between controlled laboratory experiments and real-world usage scenarios, this method highlights the significant potential of transferring underlying knowledge from high-fidelity physics-based models to data-driven models for predicting the behavior of complex dynamical systems.

**Keywords:**

lithium-ion battery; degradation diagnostics; physics-based model; hybrid-model fusion; transfer learning

# 1 Introduction

The widespread adoption of lithium-ion batteries in both electric vehicle and stationary energy storage sectors has been driven by their high energy density, decreasing costs, and extended lifespans [1-3]. However, a lingering concern within these industries revolves around the unpredictable decline in battery capacity, power, and safety over time. These factors greatly impact the performance, reliability, and safety of lithium-ion batteries. Accurate assessment of battery capacity or state of health (SOH) is crucial for mitigating premature degradation and extending battery lifespan [4]. Additionally, it aids in determining whether the battery is suitable for less demanding "second-life" applications or should be directly recycled to recover valuable constituent materials [5,6]. Hence, it is essential to develop strategies for diagnosing and prognosticating battery aging.

Many previous studies have modeled lithium-ion battery degradation by capturing aging processes, such as the formation of the solid-electrolyte interphase (SEI) layer [7,8], cracking of the active material [9,10], and lithium plating [11,12]. While such physics-based models have shown predictive success, developing models that precisely track the changes of the battery internal physical states remains challenging. Particularly, it requires profound expertise to comprehensively characterize the aging mechanism of electrochemical reactions inside the battery, and the highly nonlinear structure [13] and meticulous parameterization [14,15] of such physics-based models frequently discourage researchers in the battery community from their adoption.

An alternative approach to estimating battery SOH involves data-driven methods, which have attracted considerable attention for their ability to bypass the need for prior knowledge of the intricate physical principles governing batteries. Instead, they offer a straightforward implementation. However, data-driven methods for SOH estimation also come with limitations. They heavily rely on substantial quantities of high-quality training data [16,17], and if this data is limited or fails to represent real-world scenarios adequately, accuracy may be compromised. Additionally, models may find it difficult to effectively generalize to new conditions, and understanding the underlying reasons for predictions can be challenging, which in turn affects interpretability [18,19]. Recently, a growing body of literature [20-23] as sought to merge physics-based and data-driven models for predicting battery capacity, aiming to provide a more comprehensive and accurate depiction of complex systems. Typically, these efforts utilize physics-based models as feature extractors [23,24], incorporating features such as local stoichiometric numbers or the operational boundaries of positive and negative electrodes, or the peak and valley information of dV/dQ and dQ/dV derivatives for constant-current charging/discharging [25,26]. Nevertheless, such physics-assisted information is typically acquired during full charging/discharging cycles in laboratory environments at very low C-rates [27] and may prove less effective in practical scenarios involving moderate- and high-rate fast or partial charging strategies [28].

This paper introduces a novel hybrid fusion strategy that combines physics-based and data-driven approaches to accurately predict battery SOH. It achieves an average estimation error of only 0.63% over the entire battery lifespan using just 45 real-world data segments, together with a total of more than 1.7 million simulated data segments derived from random partial charging cycles. Leveraging a thoroughly validated physics-based battery model, we extract typical aging patterns from laboratory aging data and extend them into a more comprehensive parameter space, encompassing diverse battery aging states across different application scenarios. This simulation, based on the physics-based battery model, acts as a crucial bridge to reconcile differences between laboratory and real-world operating conditions. This approach can be widely used for rapid, high-quality data construction for data-driven models and support diagnosis and prognosis for complex systems through transfer learning in real-world scenarios.

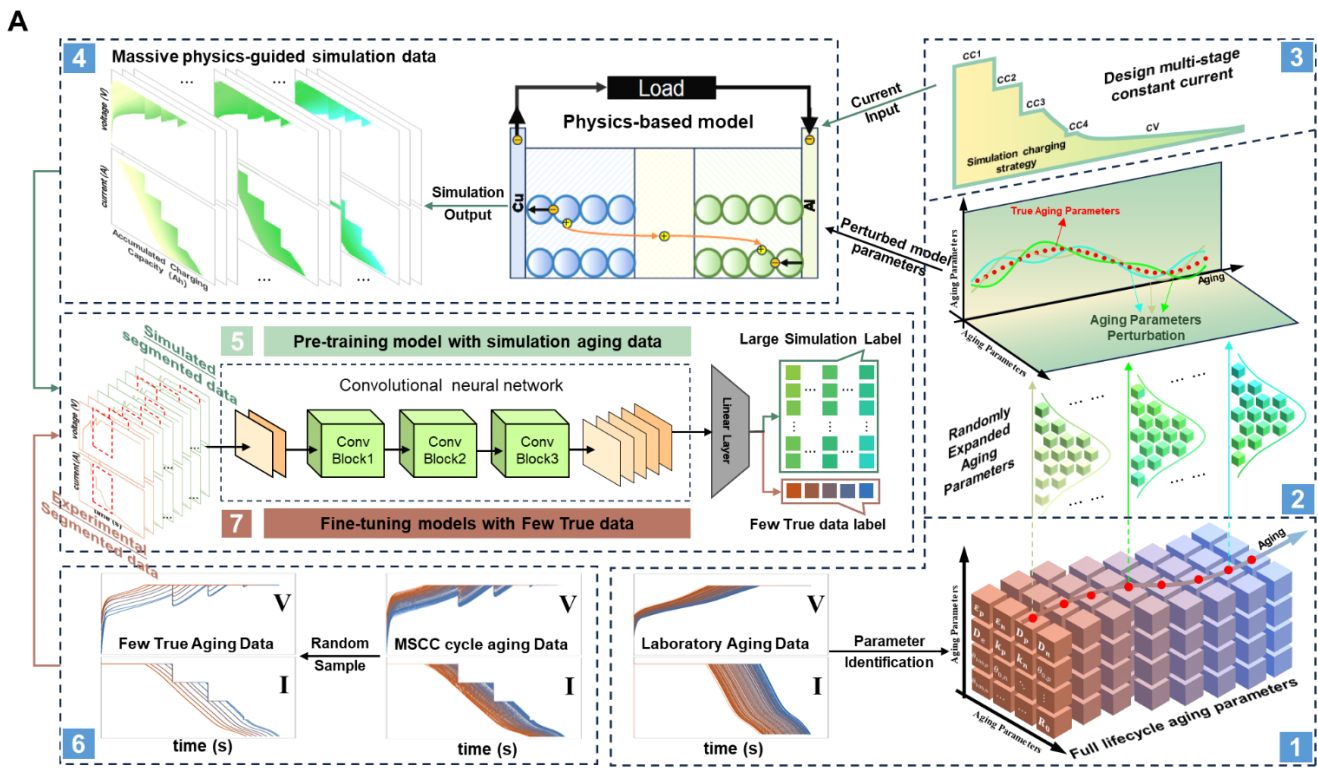

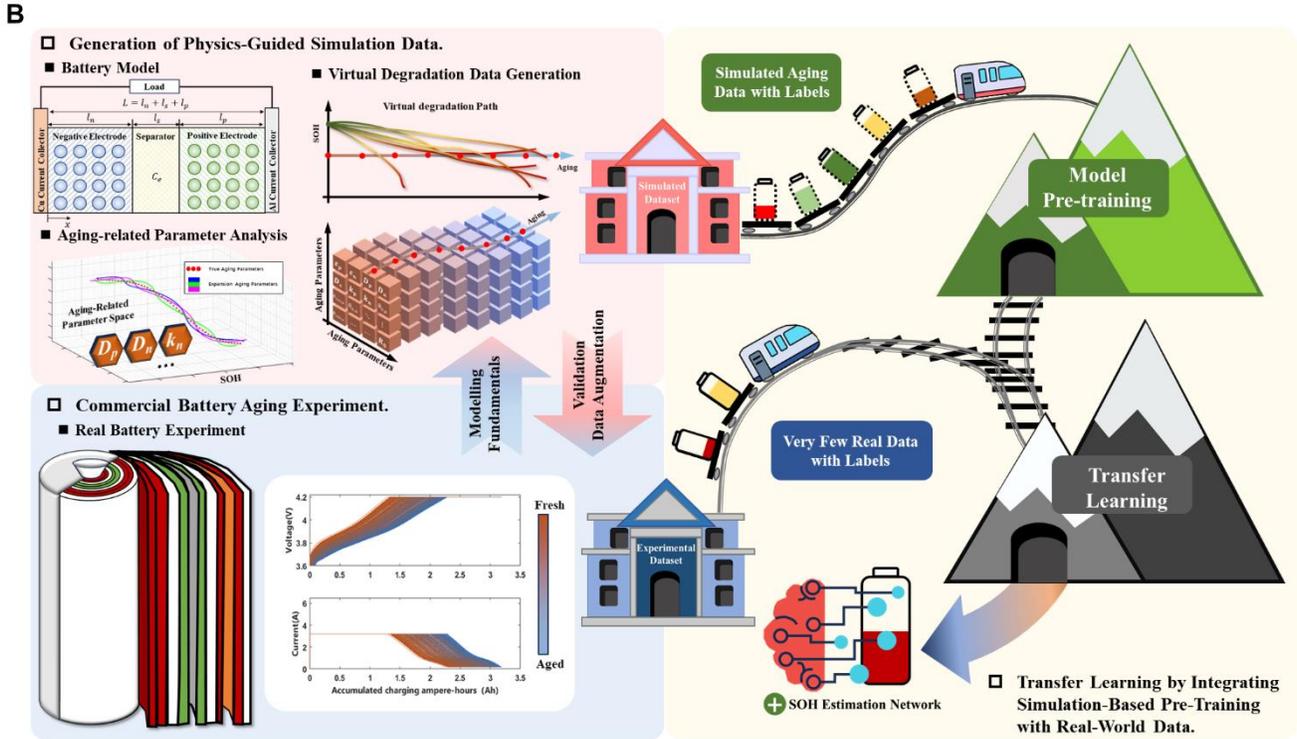

Figure 1. Schematic diagram of the proposed hybrid model fusion approach. (A) Seven steps of the proposed approach. (B) Transfer learning from physics-guided simulation data to experimental data.

Figure 1A illustrates our proposed hybrid model fusion framework. The main steps of the proposed fusion framework are summarized below:

1. Identify physical aging model parameters using laboratory aging data.

2. Expand the aging parameter space by perturbing parameters to account for cell-to-cell variations.

3. Generate real-world input current profiles and integrate them into the physics-based battery model.

4. Produce extensive physics-guided simulation data using: a) the validated physics-based battery model, b) the expanded aging parameter space, and c) input current profiles from real-world charging strategies.

5. Develop pre-training data-driven models based on physics-guided simulated aging data.

6. Fuse massive segmented simulation data with a portion of segmented experimental data from the target battery.

7. Estimate battery capacity by applying transfer learning from the pre-trained model to the

target domain.

The physics-based battery model serves as a preliminary data processor and generator, extracting hidden battery aging patterns from laboratory cycling data. This is accomplished by expanding the aging parameter space and incorporating target operating conditions, creating a bridge for transferring information from laboratory cycling experiments to real-world battery usage scenarios. This approach is also highly practical, as it requires only partial data segments of the battery with arbitrary initial SOC and degradation levels. By leveraging pre-training and transfer learning, the underlying aging information from generated and real-world data can be effectively fused, enabling accurate SOH estimation throughout the battery's life, even with merely a small amount of real data samples.

Figure 1B highlights the essence of the hybrid model fusion methodology. Using a well-verified model grounded in first-principles, we identify common aging trends from experimental data and expand these patterns into a broader parameter range, covering a variety of battery aging conditions in different practical situations. This simulation, grounded in the physics-based battery model, serves as an essential link to harmonize the discrepancies between laboratory settings and real-world usage environments. Next, a pre-training model is trained using massive high-accuracy, physics-guided simulation data. Finally, even with a minimal amount of real-world data, the foundational insights from high-fidelity models are transferred to the data-driven model for diagnostics of the lithium-ion batteries.

## 2  Experimental setup

In this work, we conducted two types of long-term cycle aging tests on a commercial INR-18650F1L battery with a nominal capacity of 3.35Ah, using three batteries for each test. The first type of experiment simulates the accelerated aging test of batteries under laboratory conditions. In this experiment, the batteries were cycled at the same discharging rate of 1C and three different charging rates of 1C, 1.5C, and 2C, to simulate the different degradation behaviors of batteries. All the cycle test was performed with a constant ambient temperature of 25℃ and charged via the commonly used constant current-constant voltage (CCCV) protocol with an upper voltage limit of 4.2V and a cutoff current of 0.16A. Following the charging step, the batteries were then discharged at 1C until the voltage drops to the lower cutoff voltage of 2.5V. The measurement data is displayed in Fig. 2A.

Compared to the simple CCCV charging condition, the second experiment focuses on the complex charging profiles typical of actual application scenarios, specifically using the multi-stage constant current (MSCC) charging strategy, increasingly prevalent for electric vehicle charging today. In the low SOC stage, the battery can safely withstand a high charging current due to the higher potential level at the anode. However, as the SOC and the degree of lithiation of the anode increase, it is necessary to decrease the charging current to reduce the polarization overpotential, avoid lithium plating, and minimize the growth of the solid electrolyte interphase (SEI) film. Therefore, the MSCC strategy employs a series of decreasing charge rates instead of a constant one, which accelerates the charging time while ensuring the battery's longevity.

In this context, an MSCC charging profile was designed for the battery aging experiment, as shown in Fig. 2B. At the start of the charging process, the charging rate was initially set to 2C, then switched to 1.5C and 1C when the SOC reached 60% and 80%, respectively. Subsequently, the battery was charged at 0.5C to the upper cutoff voltage of 4.2V and then continued charging at 4.2V until the current decayed to the cutoff value of 0.16A. Considering the increase in ohmic and polarization impedance, the battery may reach the upper cutoff voltage earlier during the constant current charging step. Therefore, at any charging rate, the battery switches to constant voltage charging mode after reaching 4.2V until the next current rate step. The cycling data for the three experimental batteries are displayed in Fig. 2C.

In this study, the data from the first type of experiment are classified as laboratory data. Based on this data, we analyze the battery's degradation behavior and extract prior knowledge through a physics-based model of the battery. The second type of experimental data simulates the battery's charging response under the MSCC charging strategy and at various aging states. This data will be utilized to evaluate the application of our SOH estimation method in real-world battery usage scenarios.

A

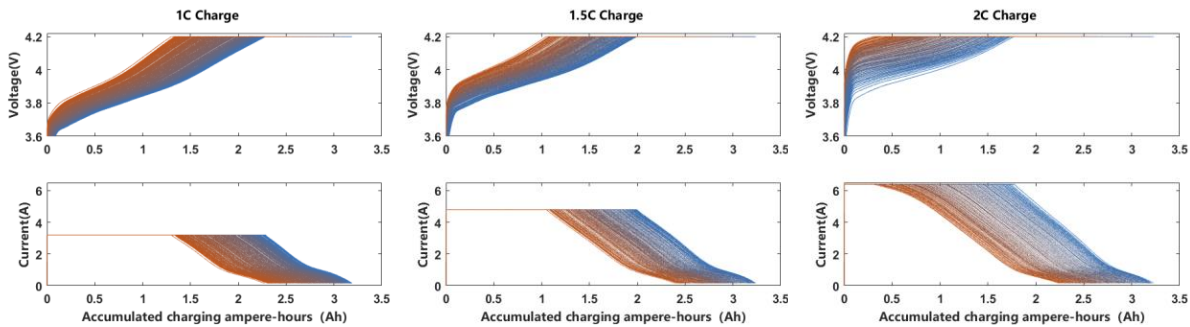

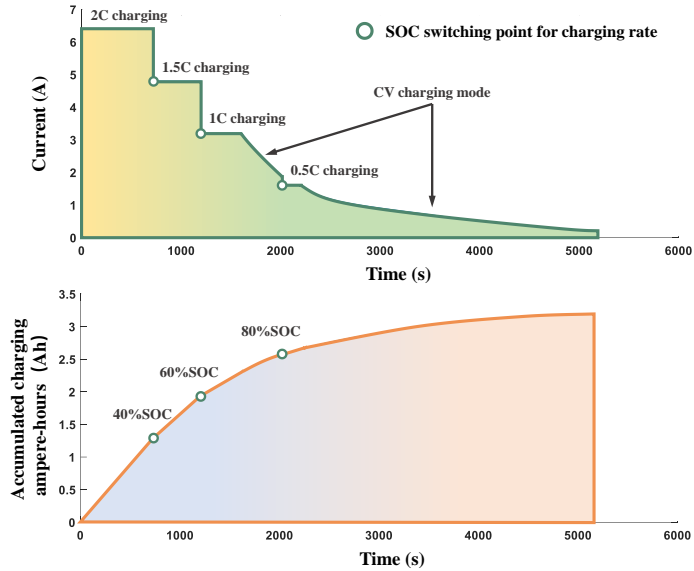

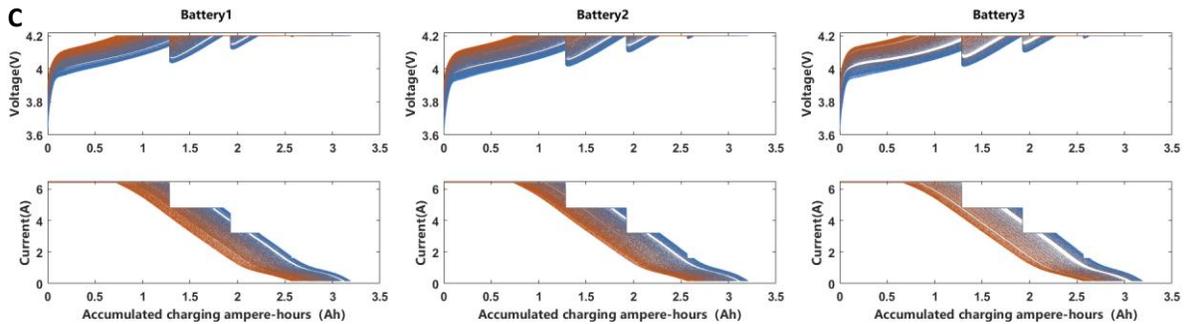

Figure 2. Two types of long-term cycle aging tests conducted in this study. (A) Cycling experimental data under single rate charging strategy. (B) Charging map of the MSCC charging strategy. (C) Cycling experimental data under MSCC charging strategy

## 3  Development of the physics-based model and parameterization

The physics-based battery model developed in this study is an extended single particle model (ESPM), which is based on the concentrated solution and porous electrode theory, integrated with electrochemical kinetics to describe the internal electrochemical behavior of the lithium battery along the electrode and separator thickness and the electrode particle radius. Figure 3 illustrates the schematic of the physics-based model and the corresponding block diagram.

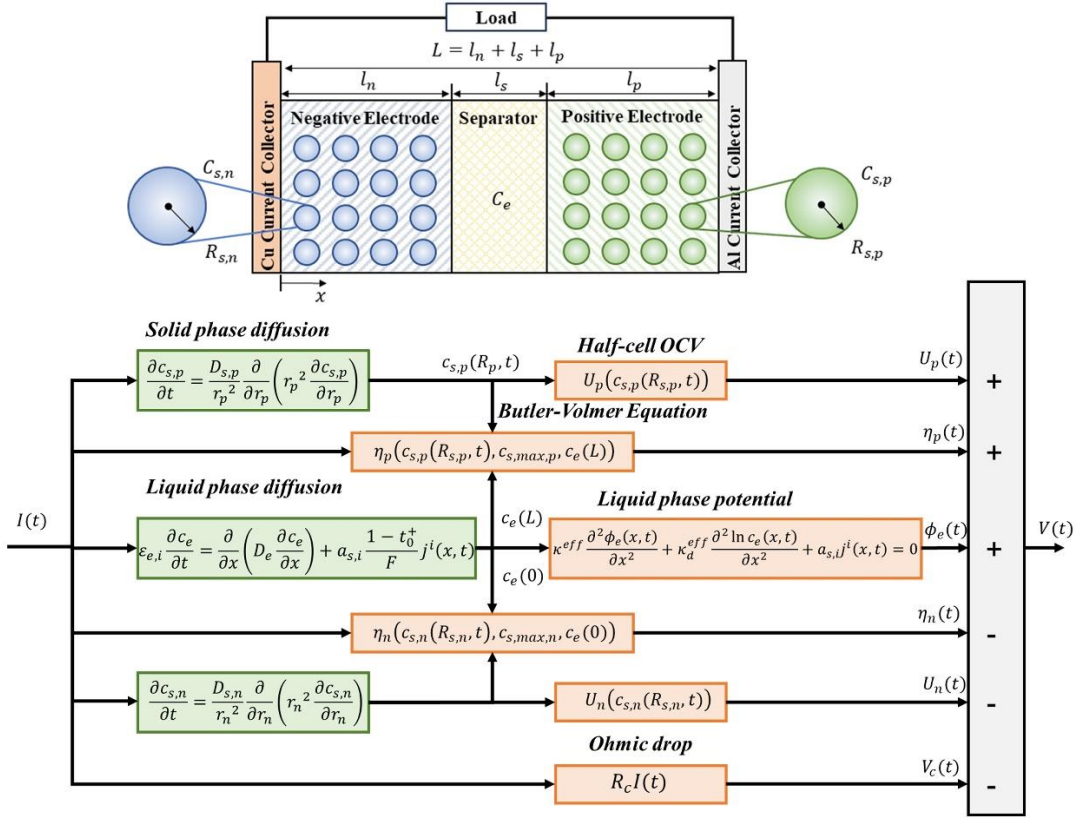

Figure 3. Schematic and block diagram of the physics-based battery model.

Originally, the model's calculation framework consisted of a series of coupled partial differential equations, making its solution complex and time-consuming. To reduce computational complexity while maintaining accuracy, we introduced several model order-reduction and approximation methods, such as the Padé approximation and polynomial approximation. Ultimately, the entire model, shown in Fig. 3, was transformed into a nonlinear state-space system, enhancing its practicality for efficient parameter identification in subsequent tasks. The derivation and simplification process of the reduced-order model is summarized below. Further details about the model can also be found Supplementary Note 1 and our previous work [4,14,32].

**3.1 Parameterization and validation of the physics-based model**

**1) Parameterization for pristine batteries**. As the physics-based battery model serve as one of the key components of simulated data generation and the hybrid fusion strategy, we first identify model parameters and validated the identified model extensively. To ensure accuracy and robustness in the identification process, experimental data spanning different current rates, ranging from 0.5C to 2C, were employed as combined identification targets, facilitating overall optimal parameter estimation.

The open-circuit voltage (OCV) of the battery reflects the difference in the equilibrium open-circuit potentials (OCP) between the cathode and anode. To decouple the two OCP curves, we conducted characterization tests on half-cells to extract the electric potentials of the positive and negative materials at different electrode lithiation degrees. The experiment used a small charge and discharge rate of 0.05C to effectively suppress polarization effects, approximating the real equilibrium potential using the average charge-discharge voltage. As a rocking-chair battery, lithium ions move back and forth between the positive and negative electrodes. The maximum usable capacities of the electrodes (in ampere hours) can be calculated as follow:

$$Q = AFL_p \varepsilon_{s,p} (\theta_{p,100\%} - \theta_{p,0\%}) c_{max,p}/3600 = AFL_n \varepsilon_{s,n} (\theta_{n,100\%} - \theta_{n,0\%}) c_{max,n}/3600 \quad (1)$$

where $Q$ is the capacity of the battery, $F$ is Faraday's constant, $\varepsilon_s$ is the volume fraction of the active material. Here, the electrochemical stoichiometric number $\theta$ is defined as the ratio of lithium-ion concentration to the maximum lithium-ion concentration of the electrode material. $\theta_{p,100\%}$, $\theta_{p,0\%}$, $\theta_{n,100\%}$, and $\theta_{n,0\%}$ are the local stoichiometric numbers or the electrode operating boundaries, as shown in Fig. 4. Those local stoichiometric numbers were determined by solving an optimization problem, where the cost function is defined as the root mean square error (RMSE) between the simulated OCV (difference between positive and negative equilibrium potential) and the measured OCV.

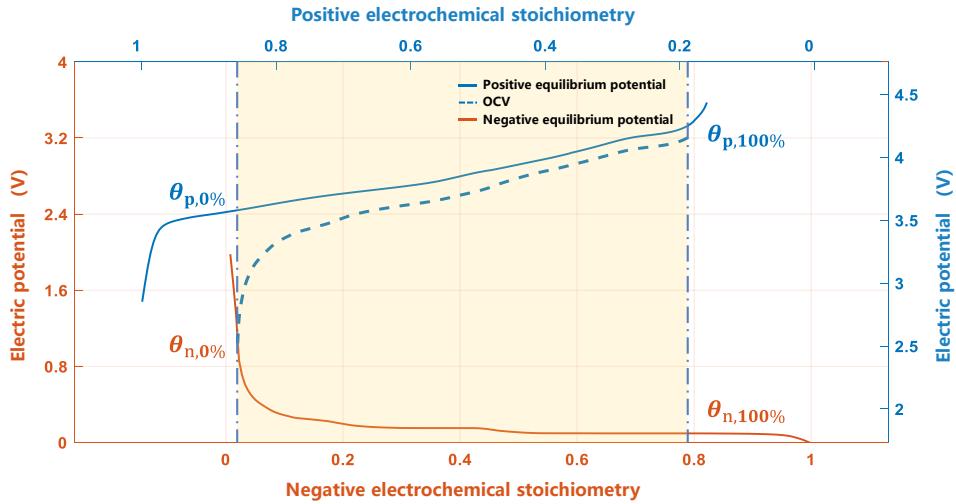

Figure 4. Correspondence between the positive and negative equilibrium potential and the OCV

The remaining model parameters were estimated by using the Adaptive Particle Swarm Optimization (APSO) algorithm. Compared to traditional algorithms such as gradient descent algorithm, APSO can search the parameter space more effectively and mitigating the risks of local optima. Moreover, the search direction of the parameters relies solely on historical optimization results, eliminating the need for additional gradient information. The update equations for the position and velocity of the particles are as follows:

$$\begin{cases} x_{n,k} = x_{n,k-1} + V_{n,k-1} \\ V_{n,k} = V_{n,k-1} * w + c_1 r_1(P_{n,best} - x_{n,k}) + c_2 r_2(P_{group,best} - x_{n,k}) \\ w_{n,k} = \begin{cases} w_{min} + (w_{max} - w_{min})\dfrac{f(x_{n,k}) - f_{min}}{f_{average} - f_{min}}, f(x_{n,k}) \le f_{average} \\ w_{max}, f(x_{n,k}) > f_{average} \end{cases} \end{cases} \quad (2)$$

where $k$ represents different iterations, $n$ represents different particle indices, $c_1$ and $c_2$ are fixed learning factors, $r_1$ and $r_2$ are random numbers in the range [0,1] with the same dimensions as the parameters to be optimized, $w$ is the inertia weight. The higher the value of $w$, the more the algorithm is inclined to explore the unknown parameter space, while a lower value makes it focus more on the neighborhood of the known optimal solution.

In each iteration, $f(x_{n,k})$ represents the fitness of the individual particle $x_{n,k}$, while $f_{min}$ and $f_{average}$ represent the minimum fitness and average fitness of all particles in the current iteration step. By adaptively adjusting the inertia weight based on the fitness distribution, the algorithm can effectively balance the global search and local search capabilities, thereby approximating the optimal point more efficiently. In this study, $w_{min}$ and $w_{max}$ are set to 0.4 and 0.9, the number of particles is set to 100, and the RMSE between the simulated voltage and measured voltage is used as the fitness function.

To ensure precision and robustness in the identification process, experimental data spanning different current rates, ranging from 0.5C to 2C, were employed as combined identification targets, facilitating overall optimal parameter estimation. The results encompassing all parameters and their respective sources are consolidated in Table 1.

To validate the generalizability of the identified parameter set, the model underwent further testing under an additional non-identification operating condition, specifically the Urban Dynamometer Driving Schedule (UDDS) dynamic condition. Comparative analysis between simulated and measured voltages is illustrated in Fig. 5, while the RMSEs for each operating condition are listed in Table 2.

Table 1. Results of the identified parameters

| Parameter | Anode | Separator | Cathode | Methods |
| --- | --- | --- | --- | --- |
| Electrode thickness (m) | 7.5e-5 | 1.2e-5 | 8.5e-5 | Destructive measurement |
| Electrode plate area ($m^2$) | 0.0802 | / | 0.0802 | Destructive measurement |
| Particle radius (m) | 5.22e-6 | / | 5.86e-6 | Destructive measurement |
| Electrolyte volume fraction | 0.31 | 0.45 | 0.26 | Liquid-liquid extraction |
| Active material volume fraction | 0.4882 | / | 0.6053 | Parameter identification |
| Maximum solid concentration ($mol/m^3$) | 58114 | / | 44871 | Parameter identification |
| Open-circuit potential (V) | $U_{ocp,neg}(\theta_n)$ | / | $U_{ocp,pos}(\theta_p)$ | Pseudo OCP measurements on half cells |
| Electrochemical stoichiometry at 0% SOC | 0.0214 | / | 0.9377 | Parameter identification |
| Electrochemical stoichiometry at 100% SOC | 0.7174 | / | 0.2717 | Parameter identification |
| Conductivity of electrolyte (S/m) | 0.963 | 0.963 | 0.963 | Experimental calibration |
| Equivalent ohmic internal resistance ($\Omega$) | 1.501e-3 | 1.501e-3 | 1.501e-3 | Parameter identification |
| Electrolyte's initial concentration ($mol/m^3$) | 1000 | 1000 | 1000 | Literature reference [14] |
| Electrolyte $Li^+$ diffusion ($m^2/s$) | 1e-9 | 1e-9 | 1e-9 | Parameter identification |
| Solid phase Li+ diffusion ($m^2/s$) | 2.787e-14 | / | 2.2006e-14 | Parameter identification |
| Reaction rate constant ($m^{2.5}/mol^{0.5}/s$) | 1.5428e-3 | / | 2.3284e-6 | Parameter identification |

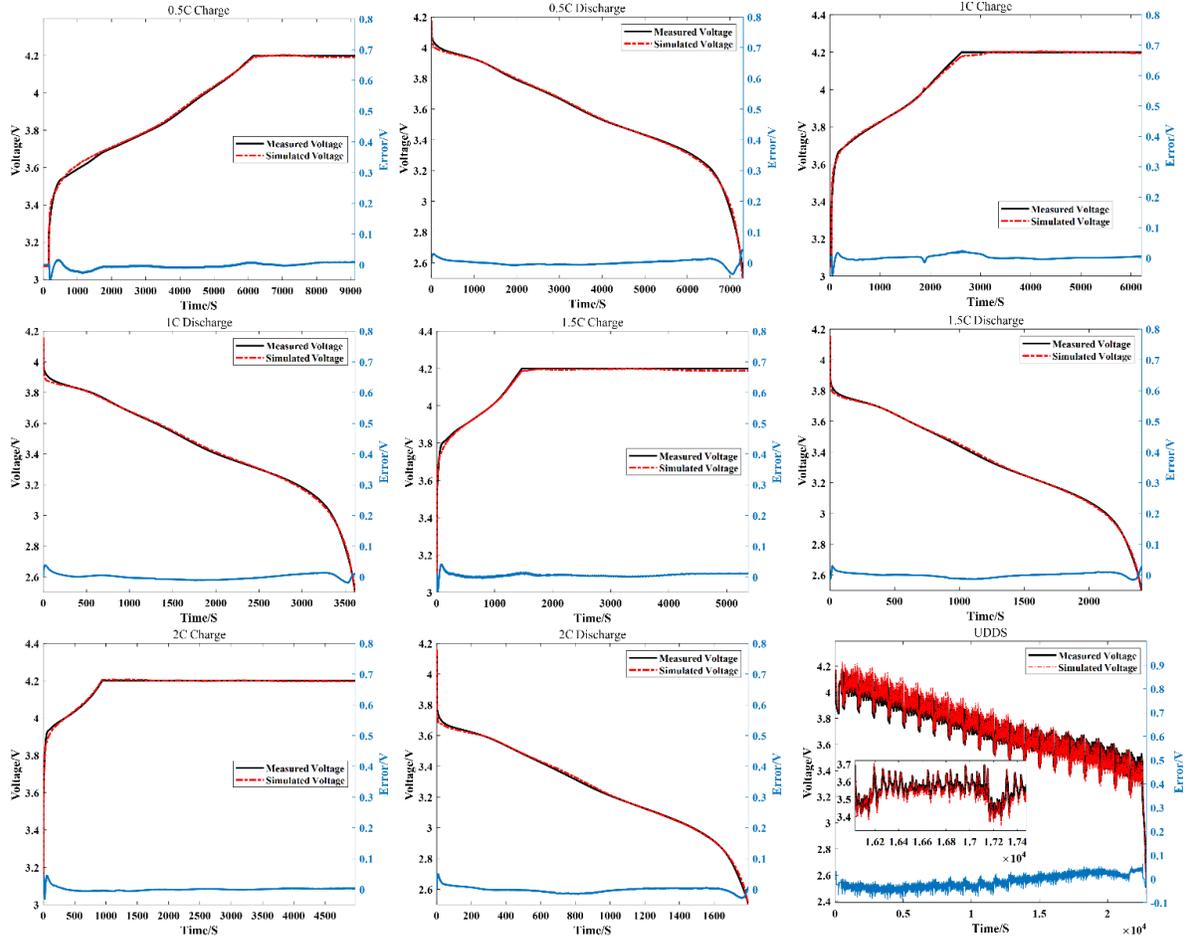

Figure 5. Comparison of simulated voltage and measured voltage under different operation conditions.

Table 2. RMSE of the reduced-ordered electrochemical battery model simulation results.

| Operating Condition | RMSE (mV) | Operating Condition | RMSE (mV) | Operating Condition | RMSE (mV) |
| --- | --- | --- | --- | --- | --- |
| 0.5C Charge | 7.81 | 1.0C Discharge | 10.33 | 2.0C Charge | 11.37 |
| 0.5C Discharge | 6.93 | 1.5C Charge | 9.73 | 2.0C Discharge | 11.06 |
| 1.0C Charge | 8.62 | 1.5C Discharge | 10.28 | UDDS | 16.34 |

The results above demonstrate the effectiveness of the physics-based model and the proposed parameter identification methodology. The developed model shows accurate voltage estimation on both the identification and test data. Furthermore, the model also demonstrates its generality by precisely simulating the battery voltage response under multiple static and dynamic operating conditions with one single set of parameters.

**2) Parameterization for batteries throughout lifespan.** To leverage aging insights gained from laboratory aging experiments, covering different degradation paths and different aging states, we then constructed a comprehensive aging parameter space through continuous parameter identification spanning the entire lifespan of the battery. From the parameters listed in Table 1 identified under pristine conditions, we selected seven aging parameters to identify their evolution within the physics-based model, tracking the battery's degradation performance throughout its lifespan. Specifically, these parameters include the volume fraction of the active material $\varepsilon_{s,p}$ and $\varepsilon_{s,n}$ in both the positive and negative electrode, the solid-state diffusion coefficient $D_{s,p}$ and $D_{s,n}$, and reaction rate kinetic constant $k_p$ and $k_n$, and the ohmic internal resistance $R_0$. These selections are pivotal for characterizing capacity fade, power fade of the performance degradation in batteries [28-31].

The laboratory aging experiments in our study involved three different charging rates (1C, 1.5C, and 2C), totaling 368 battery cycles, which is equivalent to 368 parameter combinations. Figure 6A shows the parameter identification results based on the proposed parameterization methodology. Further details about the methodology can also be found in our previous work [4,14,32]. For each experiment, we selected simulation and experimental data under six cycles across the throughout the battery's lifespan for comparison. It can be also observed in Table 3 that the physics-based battery model, together with the parameterization method could accurately predict the battery aging performance over the battery's lifetime at different conditions.

Table 3. Validation of the parameterized model under different cycles and charging rates

| 1C | | 1.5C | | 2C | |
|---|---|---|---|---|---|
| Cycle numble/SOH(%) | RMSE of voltage (mV) | Cycle numble/SOH(%) | RMSE of voltage (mV) | Cycle numble/SOH(%) | RMSE of voltage (mV) |
| 34/98.38% | 2.53 | 16/99.09% | 2.55 | 13/97.80% | 9.43 |
| 67/96.26% | 2.82 | 31/94.20% | 3.18 | 25/92.91% | 7.31 |
| 100/92.01% | 2.91 | 46/90.04% | 3.49 | 37/86.90% | 5.16 |
| 133/84.49% | 2.66 | 61/86.55% | 3.82 | 49/81.83% | 5.23 |
| 166/78.03% | 3.05 | 76/81.69% | 4.22 | 61/76.92% | 7.25 |
| 199/71.40% | 4.43 | 91/77.74% | 4.99 | 73/71.12% | 6.28 |

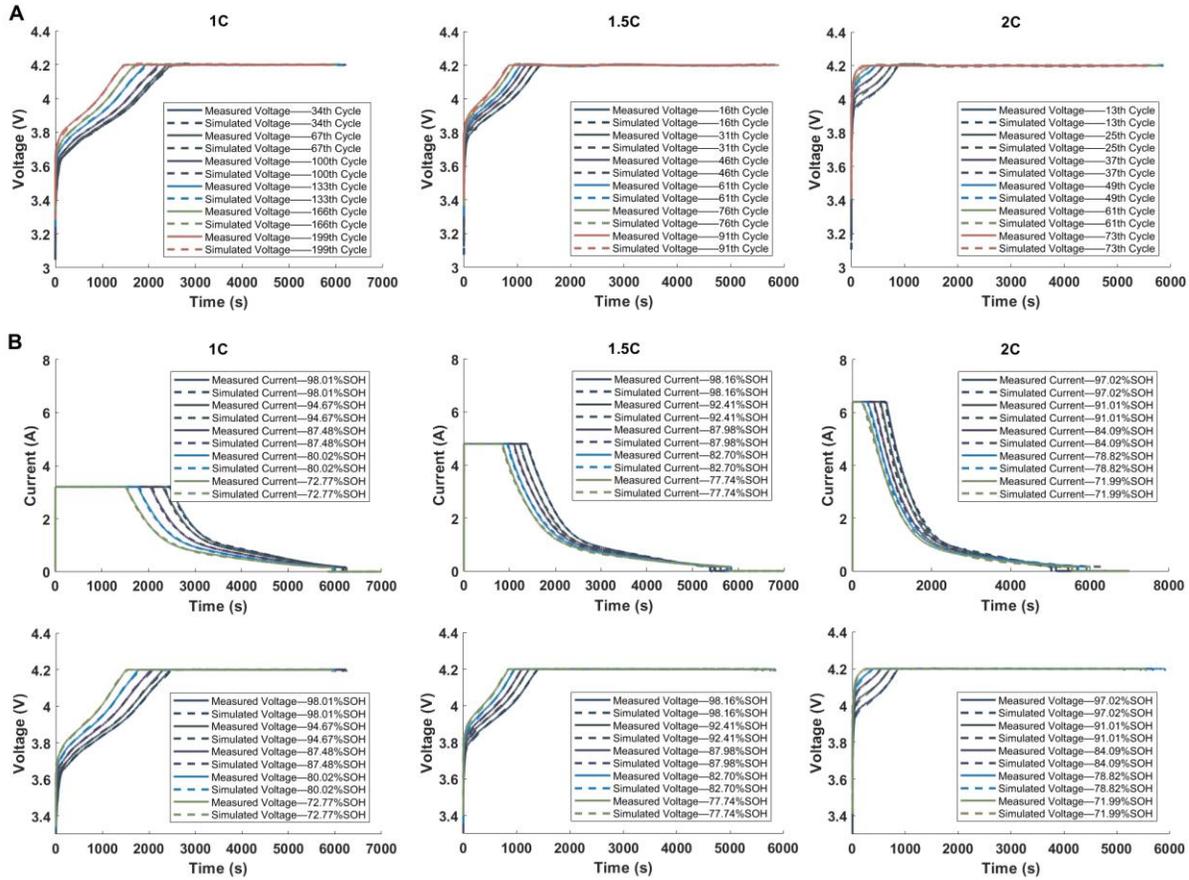

Figure 6. Parameter identification and validation of the physics-based model at laboratory conditions. (A) Model parameterization at different cycles and charging rates. (B) Validation of the parameterized model by comparing it to measured voltage and current across different cycles and charging rates.

Before employing the validated physics-based battery model for data generation and transfer learning, another critical challenge is to enable the proposed methodology with the flexibility to generate real-world input current profiles. This involves adapting the amplitude of the input current according to varying charging rates and accommodating the shortened charging times as the battery ages. Additionally, the input current profile must seamlessly transition to a constant voltage (CV) charging phase once the upper voltage limit is reached. Supplementary Note 2 documents the technical details to generate these real-world input current profiles.

Once flexible input current profiles are obtained, they can be integrated with the validated physics-based battery model to generate high-accuracy, physics-guided simulation data. Figure 6B offers selected snapshots of the physics-based model's performance throughout the battery's lifespan, demonstrating the accuracy of simulated current and voltage under various aging conditions and charging rates. It is worth mentioning that during the constant voltage (CV) charging stage, where the voltage remains stable and insensitive to battery state variations, accurately simulating the current profile is essential to enhance the performance of the upcoming data-driven model. In such cases, the data-driven model is expected to prioritize the current profile to extract key SOH features.

## 4 Development of the pre-training model

This paper expands on our previous work [33] by exploring the benefits of model-based data generation methods for data-driven models. Consequently, we adopted the optimized network structure from the previous work, as shown in Fig. 7, to further investigate model fusion methods. Table 4 lists the detailed network structure of the pre-training model. Below are the settings and global training parameters employed during the training phase: the Adaptive Moment Estimation (Adam) optimization algorithm, a batch sample size of 1024, an initial learning rate set at 0.0005, a maximum iteration limit of 30, and a learning rate decay coefficient of 5%.

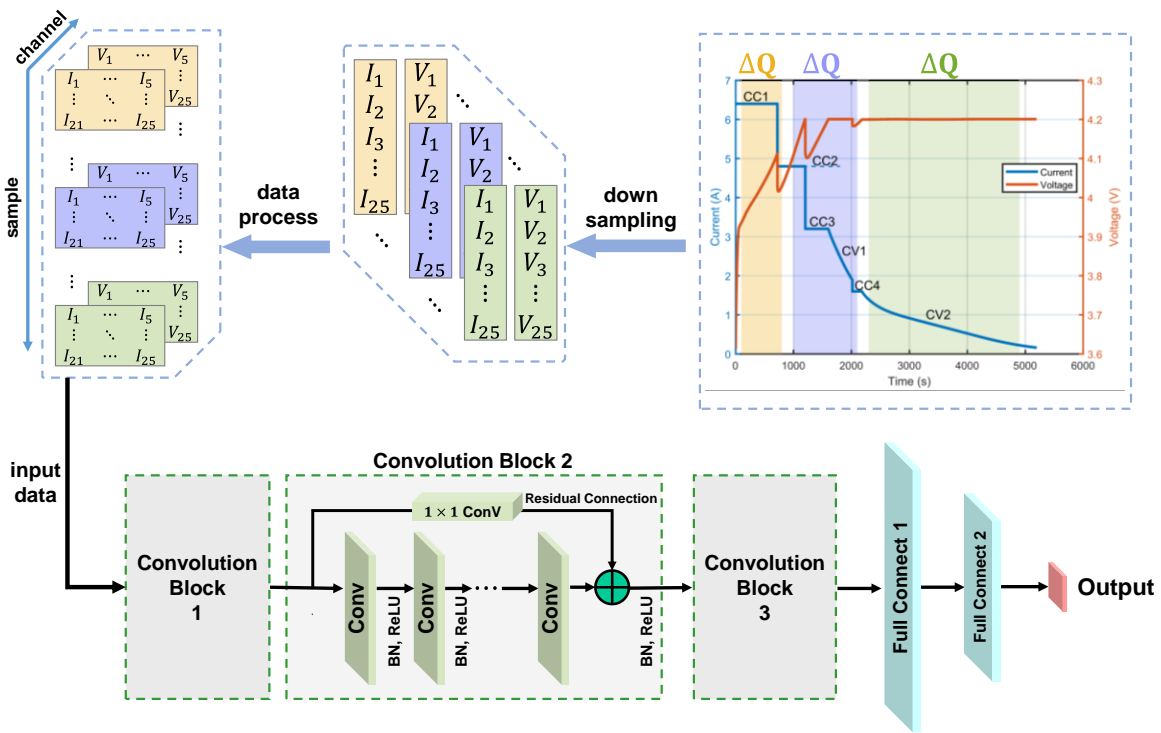

Figure 7. Schematic of input data processing and pre-training model structure

Table 4. Structure of the network for pre-training

| Layers | Sub-structure | | Output shape | Activation |
|---|---|---|---|---|
| Conv block 1 | 3 Conv Layers | 11-(2*1*2)<br>7-(2*1*11)<br>8-(2*1*7) | 5*5*8 | BN+RELU |
| Conv block 2 | 3 Conv Layers | 14-(2*3*8)<br>11-(2*3*14)<br>11-(2*3*11) | 5*5*11 | BN+RELU |
| Conv block 3 | 3 Conv Layers | 12-(5*4*11)<br>15-(5*4*12)<br>4-(5*4*15) | 5*5*4 | BN+RELU |
| FC layer 1 | neurons | 22 | 22*1 | BN+RELU |
| FC layer 2 | neurons | 74 | 77*1 | BN+RELU |

The generated data is shown in Fig. 8A, while the comparison between the measured SOH and SOH estimation results based on the pre-training model is illustrated in Fig. 8B. It's noted that all simulated data strictly adhered to the MSCC charging strategy outlined previously. During the simulated CV charging phase, the voltage was effectively maintained around 4.2V, showcasing the efficacy of the input generation method of the flexible current profile. The colors in Fig. 8A signify various battery aging states within the generated dataset, encompassing a broad spectrum. With those physics-guided simulation data, the pre-training model is expected to autonomously learn the SOH-related features at different aging stages. Figure 8B and 8C show the capacity estimation results of the pre-training model and a histogram of the estimation error. Utilizing the simulated dataset, the model achieved root-mean-square estimation error of just 0.0283Ah, equivalent to 0.84% of the nominal capacity, on the training samples. Since the simulated dataset was injected into the charging strategy tailored for the target battery, the distribution of data closely approximated the actual charging data of the battery. Consequently, the pre-training model, provided by an ample supply of simulated samples, theoretically exhibits robust generalization performance.

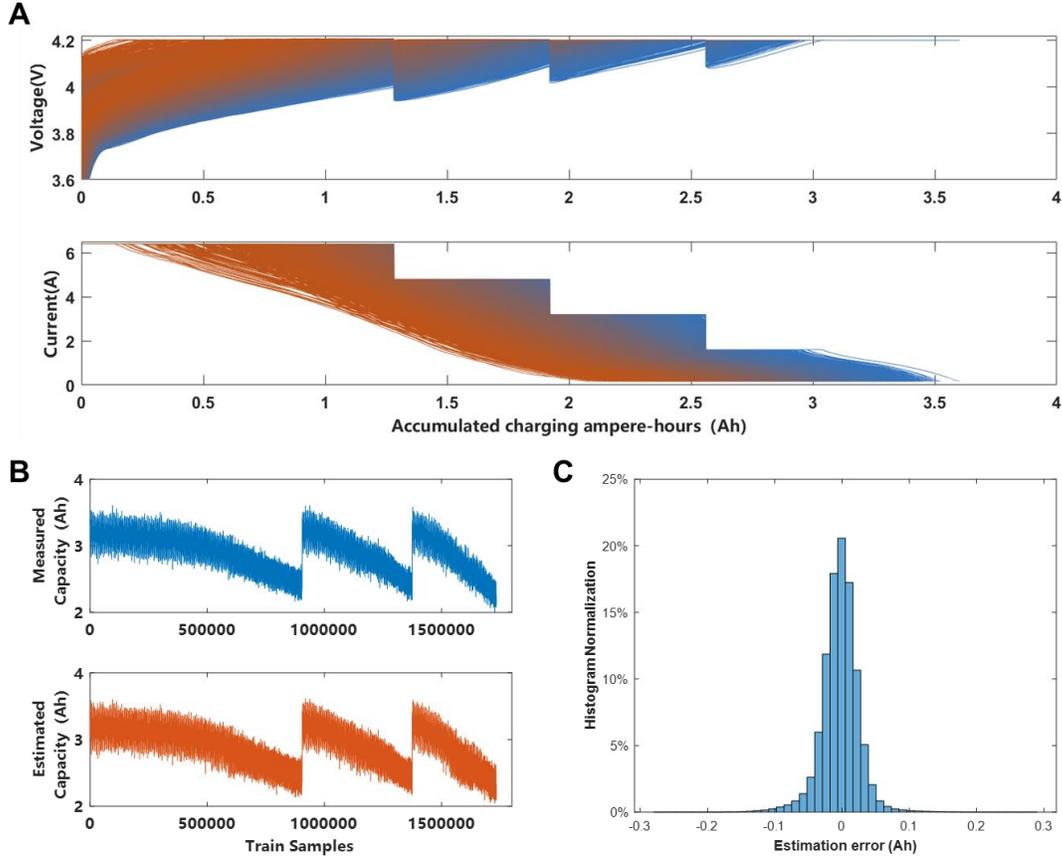

Figure 8. SOH estimation based on physics-guided simulation data (A) Large-scale simulated dataset with MSCC charging strategy (B) Capacity estimation results of the pre-training model

## 5 Results and discussion

### 5.1 Pre-training based on physics-guided simulation data for real-world scenarios

To address the complex and varied aging behaviors of batteries in practical scenarios, such as fast charging, cell-to-cell inconsistencies, and to compensate for potential issues of inadequate data for data-driven models, we applied random perturbations to the aging parameter groups. This approach enabled us to generate extensive, high-accuracy, physics-guided simulation data. Given the differing magnitudes of the aging parameters, perturbations were added proportionally to the seven aforementioned parameters as follows:

$$\begin{aligned} \text{Param}_{j,e1} &= \text{Param}_j * X_{e1} & (\ X_{e1} \sim N(1, 0.05^2)\ ) \\ \text{Param}_{j,e2} &= \text{Param}_j * X_{e2} & (\ X_{e2} \sim N(0.9, 0.05^2)\ ) \\ \text{Param}_{j,e3} &= \text{Param}_j * X_{e3} & (\ X_{e3} \sim N(1.1, 0.05^2)\ ) \end{aligned} \qquad (3)$$

The perturbation scheme involved three sets of random variables ($X_{e1}$, $X_{e2}$, $X_{e3}$), each following different normal distributions, totaling 30 random variables. The standard deviation for these distributions was set to 0.05, with different means to account for potential fixed deviations between real-world and laboratory aging parameters. This approach generated a more comprehensive and extended parameter space, as illustrated in Fig. 1A (Step 2).

Next, we combined the physics-based battery model and the current profile generation method with the expanded parameter space to simulate the multi-stage constant current (MSCC) charging scenarios. MSCC is a highly realistic representation of the charging protocols employed in real-world electric vehicle applications today, especially for fast charging applications. In this work, we employ the MSCC charging scenarios to demonstrate the effectiveness of our proposed model under real-world fast charging conditions.

Through the introduction of random perturbations, the aging parameter set expanded by a factor of 90, yielding 33,120 parameter combinations, satisfying the diversity requirements of the samples. Moreover, leveraging the data segmentation rule outlined in Supplementary Note 3, we further generated 1,733,782 data segments for training purposes.

**5.2 SOH estimation by hybrid model fusion and transfer learning**

Within the proposed framework for SOH estimation, a small portion of data segments is allocated for transfer learning, while the remainder is reserved for validation purposes to assess the SOH estimation performance of the transferred model. Validation is conducted across all cycles of three experimental batteries. For each cycle, five data segments with arbitrary initial SOC and a fixed capacity increment of 1.5Ah are randomly chosen. As a benchmark for evaluating model performance, Fig. 9A displays the SOH estimation results from the pre-trained model, which was trained solely based on physics-guided simulation data, but validated using all actual MSCC experimental data.

The pre-training model achieved stable SOH estimation throughout the whole life of the three target batteries. The majority of the estimation errors can be controlled within 5%, as demonstrated by the bounding region in the figure. Despite the considerable capacity recovery of Battery 1 and Battery 2 (around 70 cycles, due to a period of long-time rest during cycling test), the pre-training model can still accurately track the SOH of different batteries. This observation highlights that our model could mainly focuses on the current state of the battery, as indicated by its charging behavior, and remains unaffected by the battery's historical aging trajectory. In essence, the model demonstrates stability and insensitivity to the various aging modes encountered by the battery in real-world scenarios.

To further enhance the model's performance, we conducted transfer learning to fine-tune the embedded parameters of the data-driven model [33]. The training data is an extended dataset that combines a small amount of real data with plenty of simulated data. Based on the same segmentation rules, 45 actual data segments were randomly extracted from the arbitrary SOC and arbitrary cycle of three experimental batteries. The resulting data was then down-sampled and reconstructed into a 5×5×2 matrix to meet the requirements of the network input. After that, 50,000 samples of simulated data segments were randomly selected from the pool of 1,733,782 segments to complete the dataset for transfer learning.

In order to increase the proportion of actual data in the extended dataset and expedite the transfer of the pre-training model to the target domain, we replicated the 45 actual data segments to yield 5,000 samples and then merged them with the 50,000 pieces of simulated data. It is important to note that the replication process involves only repeated sampling and does not introduce any additional information of the actual operation data.

During the transfer process, the maximum number of iterations and initial learning rate were set to 5 and 1e-5, respectively. The computational cost of transfer learning was substantially reduced by 99.471% compared to the pre-training model, which was trained for 30 iterations using 1,733,782 simulated data samples. The SOH estimation results of the transferred model across the entire lifespan of the three target batteries are illustrated in Fig. 9B.

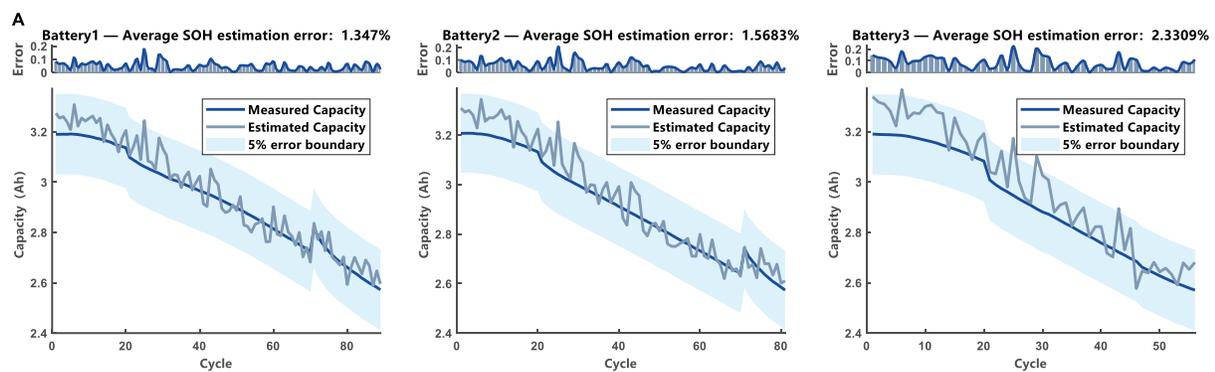

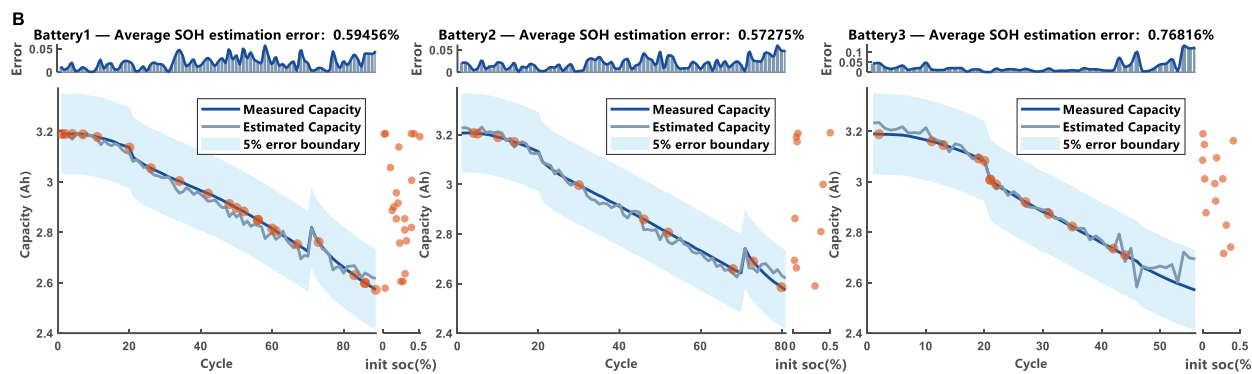

Figure 9. Capacity estimation by the pre-trained and transfer learning models. (A) Capacity estimation results from the pre-trained model, which was trained solely based on physics-guided simulation data. (B) Capacity estimation results of the transferred model by incorporating 45 actual data segments

The results demonstrate that the transfer learning model is capable of accurately estimating the SOH of the target batteries with minimal reliance on their actual operational data. In Fig. 9B, the red dots represent the distribution of actual data samples. Here, it's important to note that although there are 45 cycles marked with discrete red dots (some dots may overlap as data from the same cycle could be randomly selected more than once), each red dot represents only a random data segment extracted from the entire cycle. The information provided by a single segment is significantly less comprehensive than that of a complete cycle. For example, based on the operational data of three experimental batteries, a total of 12,946 segments can be extracted from 226 cycles according to the segmentation rules outlined in Supplementary Note 3. Among these segments, the 45 samples represented by the red dots account for only 0.3476% of the overall dataset. Even with the addition of such a small portion of actual data samples, the transfer learning model decreases the average SOH estimation errors for the three batteries from 1.35%, 1.57%, and 2.33% to 0.59%, 0.57%, and 0.77%, respectively. This signifies a substantial improvement in estimation accuracy, achieving at least a 50% enhancement with the proposed transfer learning approach.

Furthermore, it is evident that despite three experimental batteries undergoing identical cycling conditions (MSCC charging and 1C discharging protocols), they displayed varying degradation rates. This variability can be attributed to inconsistencies in battery manufacturing. Similarly, in real-world scenarios, applying identical charging and discharging currents to batteries within a single module can lead to inconsistencies during prolonged operation. Nevertheless, the proposed transfer learning model successfully captures these varying degradation rates.

To further investigate the impact of the distribution of actual data samples on model performance, we then randomly selected 45 charging segments from the datasets of Battery 1 and Battery 2, merging them with simulated data. This analysis aimed to evaluate the influence of data from certain batteries on others for which data is not yet available. The subsequent SOH estimation results of the transferred network are presented in Fig. 10A below.

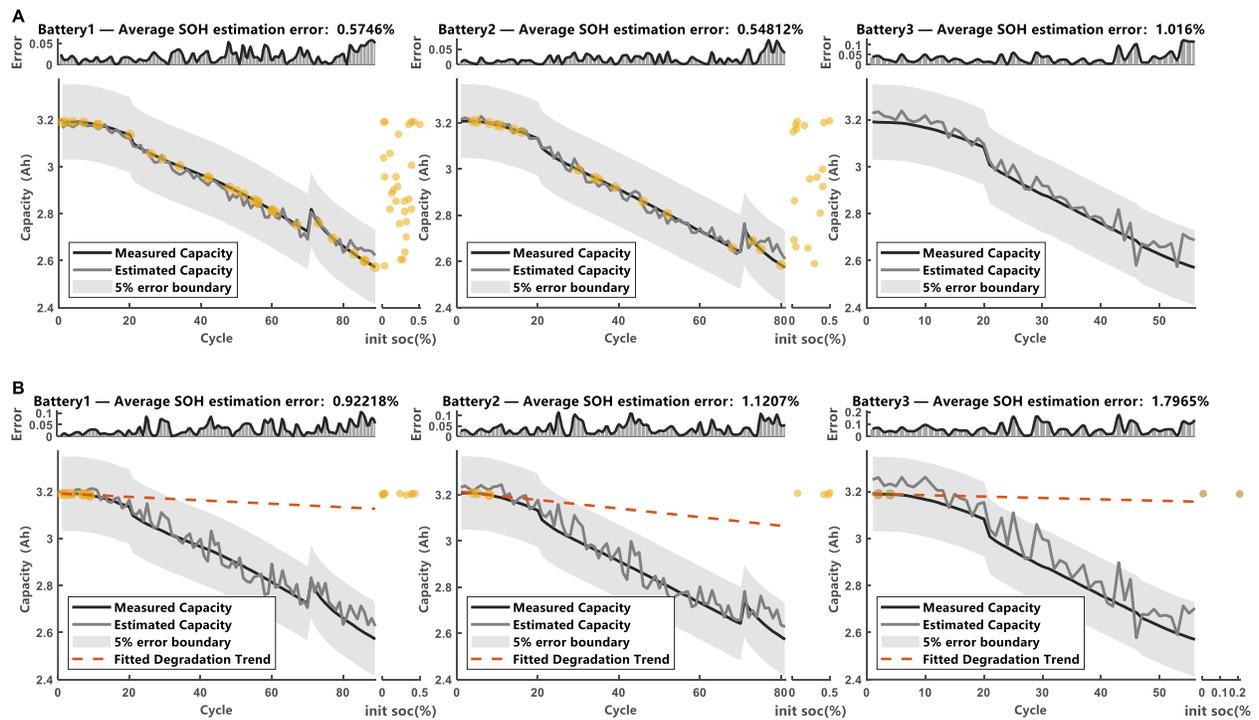

Figure 10. Robust performance with limited actual data (A) Capacity estimation results of the transferred model using 45 actual data segments only from two batteries (B) Capacity estimation results of the transferred model using 15 early-life data segments from three batteries

By comparing Fig. 9A and Fig. 10A, it is evident that the accuracy of the SOH estimation for Battery 1 and Battery 2 improved significantly due to the introduction of actual data samples. Additionally, despite no actual data sample from Battery 3 was used, its SOH estimation error also decreased from 2.3309% to 1.016%, indicating a reduction of 56.41%. This finding underscores the potential for information sharing within the pre-training model, which, established using a substantial amount of validated simulated data, can benefit from the utilization of small labeled data from partial batteries. Consequently, the transferred model demonstrates enhanced SOH estimation capability for batteries subjected to the similar operating conditions but exhibiting different aging states.

Following the discussion on the impact of actual data distribution, we also investigated the model's performance using solely early-life data. This exploration aims to evaluate the model's ability to extrapolate to moderate-life and late-life data, shedding light on its capability to accurately capture degradation characteristics across various stages of battery life.

To achieve this, we selected 15 segments from only the first ten cycles of the three batteries and added them into the training dataset after replication. This configuration is devised to guide the model in learning precise degradation characteristics from actual early-life data, while also ensuring its adaptability to moderate-life and late-life estimation by incorporating massive physics-guided simulation data. The performance of the model is illustrated in Fig. 10B, showcasing its exceptional ability to tackle the nonlinear aging problem of batteries using solely early-life data. This performance surpasses that of some other fitting classes of methods, such as the linear extrapolation (red dash lines), affirming the robustness and effectiveness of our approach.

Compared to the pre-training model, the transferred model demonstrates a notable enhancement in SOH estimation accuracy for early-life data. This improvement was anticipated as a result of incorporating corresponding actual data during transfer learning. Furthermore, comparing Fig. 9A and Fig. 10B reveals that the model's performance improves for moderate-life estimation, likely due to extrapolating degradation information from early-life data. In contrast, the accuracy of late-life estimation remains relatively stable. This could be attributed to the diminishing representativeness of the introduced information as batteries exhibit varying charging behaviors during severe aging stages. Additionally, the substantial volume of simulated data within the extended training dataset plays a role in preventing performance degradation during transfer learning, contributing to the model's consistent performance across different stages of battery life.

To further investigate the effect of the number of actual data samples on the accuracy of the transferred model, we implemented a series of comparative experiments where the sample sizes of actual data segments were set to 15, 30, 45, 60, 75, and 90, respectively. To ensure a fair comparison, all actual data segments were replicated to 5,000 samples and combined with the same 50,000 simulated data segments. The comparison results of the model accuracy are presented in Fig. 11A.

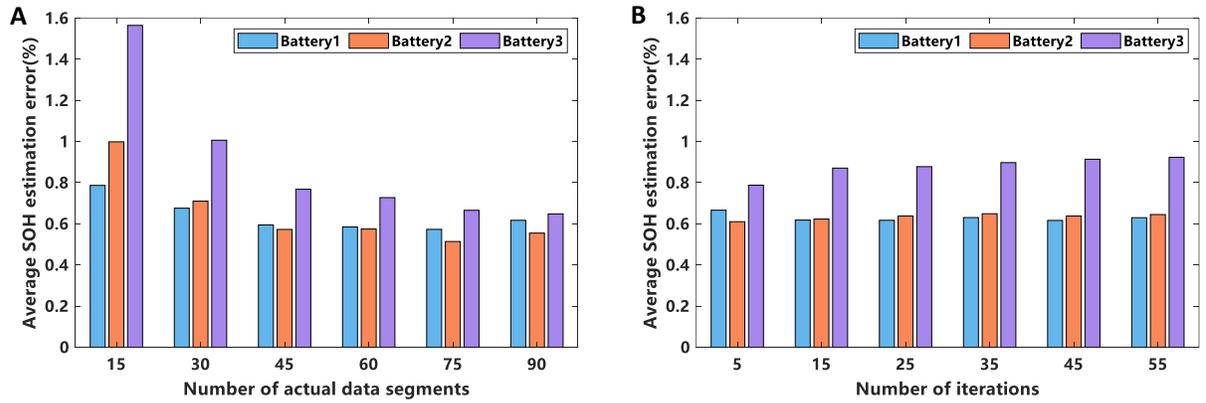

Figure 11. Impact of different factors on the accuracy of the transferred model. (A) Average estimation error of transferred models with different number of actual data segments (B) Average estimation error of transferred models with different number of iterations

The results indicate that as the number of actual data samples increases, the model accuracy gradually increases until it reaches saturation. The leftmost column shows that even with only 15 actual data segments, the average SOH estimation error on an arbitrary target battery can be controlled within 1.6%. In terms of the distribution of SOH estimation errors over all cycles of the target batteries, 87.2% of the errors are within 2%, with an overall average error of 1.1%. With an increase in the number of actual data samples to 45, the transferred model achieves an average error of 0.63% on all test data, where 97.8% of the errors are controlled within 2% and 81.9% of the errors fall within 1%.

The maximum number of iterations for transfer learning was set to 5 during all above validation. To investigate the impact of the number of iterations on model accuracy, we conducted a study with different maximum iterations ranging from 5 to 55. As illustrated in Fig. 11B, the model accuracy remains relatively stable as the number of iterations increases. This enables accurate estimation of SOH on target batteries with only a few transfer iterations. The information latent in the actual data can be effectively utilized without extensive iterative training.

# 6  Conclusions

In summary, this paper introduces a novel hybrid fusion strategy that combines physics-based and data-driven approaches to predict battery capacity with high accuracy. Achieving an average estimation error of only 0.63% over the entire battery lifespan, this method only utilizes 45 real-world data segments along with over 1.7 million simulated data segments from random partial charging cycles. By leveraging a validated physics-based battery model, the strategy extracts typical aging patterns from laboratory data and extends them into a comprehensive parameter space, encompassing diverse battery aging states and practical cell-to-cell variations. The integration of transfer learning enables the effective fusion of knowledge from high-fidelity physics-based models to data-driven models, significantly enhancing the accuracy and reliability of battery SOH estimation.

This hybrid approach has profound implications for the management of EVs and stationary energy storage systems, offering improved battery lifespan and performance. It supports the assessment of battery suitability for second-life applications or recycling, contributing to sustainability. By bridging the gap between controlled laboratory experiments and real-world usage scenarios, this method enhances the applicability of predictions in practical contexts. Potential applications span from EVs and renewable energy storage systems to battery manufacturing, quality control, and consumer electronics, promising better performance, reduced costs, and extended battery life across various industries. In general, this approach can complement methods based on purely first-principle models or data-driven models. Broadly speaking, this work highlights the promise of hybrid models in addressing complex real-world problems by leveraging the strengths of both physics-based and data-driven techniques.

## Supplementary information

Supplemental descriptions of the techniques used in this work, along with relevant analyses are detailed in supplemental information.

## Acknowledgements

This work is funded by the National Natural Science Foundation of China (Grant No. 52307246 and 52177218) and the Natural Science Foundation of Shanghai (Grant No. 23ZR1429100).

# Declaration of competing interest

Yisheng Liu, Guodong Fan and Xi Zhang have filed a patent related to this work: CN Application No. 202310842551.1, dated 11 July 2023.